\documentclass[showkeys]{revtex4}
\usepackage{epsfig}
\usepackage{amsmath,amssymb,bm}
\newcommand{\be}{\begin{equation}}
\newcommand{\ee}{\end{equation}}
\newcommand{\la}{\langle}
\newcommand{\ra}{\rangle}
\newcommand{\clDq}{{\cal D}_q}

\begin{document}

\title{\bf Time Scales in Futures Markets and Applications}

\author{Laurent Schoeffel}
\affiliation{CEA Saclay, Irfu/SPP, 91191 Gif/Yvette Cedex,  France}

\begin{abstract}
The probability distribution of log-returns for
financial time series, sampled at high frequency, 
is the basis for any further developments in quantitative finance.
In this letter, we present experimental results based on a large set of
time series on futures.
We show that the t-distribution with $\nu \simeq 3$ gives a nice description
of almost all data series considered for a time scale $\Delta t$ below $1$ hour.
For $\Delta t \ge 8$ hours, the Gaussian regime is reached.
A particular focus has been put on the DAX and Euro futures.
This appears to be
a quite general result that stays robust on a large set of futures, but not on any data sets.
In this sense, this is not universal.
A technique using factorial moments defined on 
a sequence of returns is described and similar results 
for time scales are obtained.
Let us note that from a fundamental point of view, there is no clear reason why DAX and Euro futures
should present similar behavior with respect to their return distributions.
Both are complex markets 
where many  internal and external factors interact at each instant to determine the transaction price.
These factors are certainly different for an index on a change parity (Euro) and an index on stocks (DAX).
Thus, this is striking that we can identify universal statistical features in price fluctuations
of these markets. This is really the advantage of micro-structure analysis to prompt unified approaches 
of different kinds of markets. 
Finally, we examine the relation of power law distribution of returns with another scaling behavior of the
data encoded into the Hurst exponent.
We have obtained
$H=0.54 \pm 0.04$ for DAX and $H=0.51 \pm 0.03$ for Euro futures.
\end{abstract}


\maketitle

%=========================================================================
\section{Introduction}
%=========================================================================
%\noindent

Returns in financial time series are fundamental inputs to quantitative finance.
To a certain extend, they provide some insights in the the dynamical content of the market.
There are several experimental analysis showing that the probability distributions of returns
in a large set of financial markets exhibit power law tails \cite{a1,a2,a3,a4}. 

Let us consider financial price series, labeled as $S(t)$, from which
we extract the log-returns $x(t)=\log [S(t+\Delta t)/S(t)]$ over some time interval $\Delta t$.  
Any statistical analysis can then be conducted on these
log-returns $x(t)$.
There are  evidences that 
for mature and high liquid markets, in particular futures, both positive and
negative tails conform to the so-called inverse cubic law \cite{b1,b2,b3,b4,b5,b6}.
It means that if we express the  the  distribution of returns as a power law 
\begin{equation}
\label{pl}
f(|x|) \sim \frac{1}{|x|^{\nu+1}}
\end{equation}
at large values of $|x|$, we can measure the exponent $\nu \simeq 3$ 
as it is done in Ref. \cite{b5,b6}. This is correct
for a variety of mature markets, in particular futures.
This observation is in contrast to predictions from the Pareto-Levy distribution, 
and former experimental results, from which we expect $1 < \nu < 2$ \cite{c1}.
This discrepancy in results shows how the topic is critical and a clear driver for theoretical considerations.
There is no universality and the use of large samples of high frequency data
is a key point in the understanding of the underlying market dynamics.

Very generally, all experimental investigations confirm that distribution of returns evolve from
power law behavior at small time scales $\Delta t$ (see Eq. (\ref{pl})) to Gaussian at large time scales.
The precise value of the exponent $\nu$ depends on the market under consideration with a tendency
for mature and liquid markets to fall outside the Levy stable regime, namely  $\nu>2$.

In this letter, we consider a few financial series on futures, which correspond to mature markets with high level of liquidity.
First, we discuss the universality of the experimental distributions of returns on these financial series.
Following Ref. \cite{b6}, we extract a parameterization of the log-returns distribution which confirms the previous
result, $\nu \simeq 3$, for small time scales $\Delta t$. 

Then, we show at which value of the time scale, market dynamics enters into the Gaussian regime.
We discuss the universality of this particular scale with respect to a sample of high liquid futures.
In a second part, we address the determination of these  scales using an analysis technique based on factorial moments
\cite{d1}. We prove  the consistency between the
appearance of intermittency and the deviation from the Gaussian regime, as derived in the first part.

From the measurement of the distributions for returns $x(t)=\log [S(t+\Delta t)/S(t)]$, we can understand 
how these returns sum up to build the quantity 
$x(t,\tau)=\log [S(t+\tau)/S(t)]$, where $\tau$ is a multiple of $\Delta t$.
Then, in a thirst part, we discuss the scaling of $<x(t,\tau)^2>_t$ with $\tau$.
We  relate the so-called Hurst exponent to the 
power law tail $\nu$.

%=========================================================================
\section{Distribution of Returns $P(x)[\Delta t]$}
%=========================================================================
\label{returns}

%=========================================================================
\subsection{Definitions}
%=========================================================================

From standard quantitative analysis \cite{a1,a2,a3,a4}, we know that the distribution of log-returns $x(t)$,
namely $P(x)$, can be written quite generally as
\begin{equation}
\label{ini}
P(x)=\frac{1}{Z} \exp( - \frac{2}{D} w(x)/2)
\end{equation}
where $w(x)$ is an objective function and $Z$ a normalization factor. In particular, it can be shown easily that
 $w(x)$ can be derived by minimizing a generating functional $F[w(x)]$, subject to some constraints on
the mean value of the objective function. 
In Eq. (\ref{ini}) we do not specify the time scale $\Delta t$ at which returns are derived. 
When need, we will take this variable into account with the notation $P(x)[\Delta t]$.
It reads
\begin{equation}
\label{ini2}
F = \int dx P(x) \left[ \log P(x) + w(x)/D -\lambda \right]
\end{equation}
where $\lambda$ is an arbitrary constant.

In addition, the expression given in Eq. (\ref{ini})  for the probability distribution 
can also be seen as the outcome of an equation of motion for $x(t)$. From Eq. (\ref{ini}) and (\ref{ini2}),
we can express the stochastic process $x(t)$ as 
 a Markovian  process of the form \cite{a1,a2,a3,a4}
\begin{equation}
\label{eq1}
\frac{dx}{dt} = f(x) + g(x) \epsilon(t)
\end{equation}
where $\epsilon(t)$ is a Gaussian process satisfying $<\epsilon(t)\epsilon(t')>=D \delta(t-t')$ and $<\epsilon(t)>=0$.
In Eq. (\ref{eq1}), functions $f$ and $g$ depends only on $x(t)$. Adopting the It\^o convention
\cite{a1,a2,a3,a4}, the distribution 
function $P(x,t)$, associated with this equation of motion (Eq. (\ref{eq1})), is given by the 
following Fokker-Planck equation
\begin{equation}
\label{eq2}
\frac{\partial P(x,t)}{\partial t}= 
\frac{\partial^2}{\partial x^2} [\frac{D}{2} g^2(x) P(x,t)]
-
\frac{\partial}{\partial x} [f(x) P(x,t)]
\end{equation}
From Eq. (\ref{eq2}), we can finally extract the stationary solution for $P(x)$ in the form of Eq. (\ref{ini})

\begin{equation}
\label{eq3}
P(x) = \frac{1}{Z} \exp \left[ -\frac{2}{D} \int dx \frac{D g \frac{dg}{dx} -f}{g^2}   \right]
\end{equation}

Whether Eq. (\ref{eq1}), (\ref{eq2}) and (\ref{eq3}) can be related to real data on financial markets is
not granted. Therefore, we need to compare predictions derived from these equations to real
data. As mentioned above, we use financial time series on different futures, using a five minutes sampling.

In Eq. (\ref{eq1}), (\ref{eq2}) and (\ref{eq3}), functions $f$ and $g$ are not specified and any choice can be considered.
Obviously, only specific choices will have a chance to get a reasonable agreement with real data.
For example, let us consider three cases:
\begin{itemize}

\item[(i)]
If $f(x)=-x$ and $g(x)=1$, we obtain $P(x)Z = \exp(-x^2/D)$, and thus we predict a Gaussian shape for the log-returns
distribution.
\item[(ii)]
In the more general case where $f(x)=\lambda g \frac{dg}{dx}$ and
$g$ is not constant, we obtain 
$$
P(x)Z = \frac{1}{g^{2(1-\lambda/D)}}
$$
and thus we predict non-Gaussian shape for the log-returns
distribution.
\item[(iii)] Let us specify the case (ii).
Introducing a constant $\nu$ and defining the two functions $f$ and $g$ as
$
f(x)= \frac{D(3-\nu)}{4} \frac{2x}{\nu} (1+\frac{x^2}{\nu})
$ and
$
g(x)=(1+\frac{x^2}{\nu})
$,
we get
\begin{equation}
\label{nu}
P(x)Z = \frac{1}{(1+\frac{x^2}{\nu})^{(\nu+1)/2}}.
\end{equation}
In this scenario $P(x)$ follows the so-called t-distribution.
It depends on one parameter $\nu$ to be fitted on real data, for normalized log-returns.
Let us notice that Eq. (\ref{nu}) is equivalent to the
q-exponential form of Ref. \cite{ts1,ts2,ts3}
\begin{equation}
\label{q}
P(x)Z = \frac{1}{(1+x^2 \frac{q-1}{3-q})^{1/(q-1)}}.
\end{equation}
where we have conserved the notations of Ref. \cite{ts1,ts2,ts3}.
Obviously, Eq. (\ref{nu}) and (\ref{q}) are directly related by
$(\nu+1) /2 = 1/(q-1)$. In particular, $\nu=3$ is equivalent to $q=1.5$.

\end{itemize}

\begin{figure}[htbp]
  \begin{center}
    \includegraphics[width=0.45\textwidth]{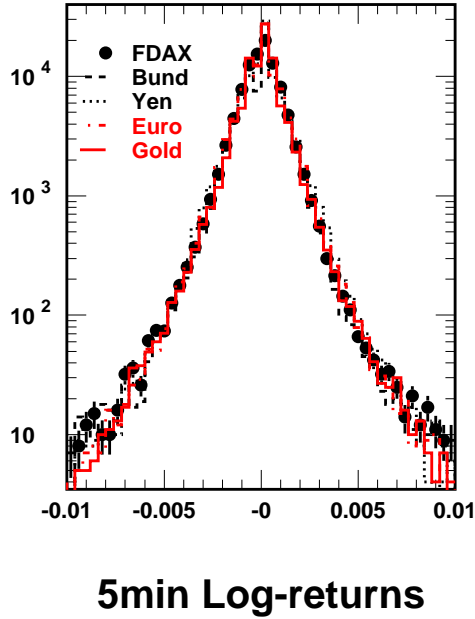}
  \end{center}
  \caption{Log-returns  $x$ (five minutes sampling) for a large set of futures.
To make the comparison, we have scaled $x$ for all futures to the volatility of the DAX futures (FDAX).
we observe that futures on DAX, Bund, Yen, Euro, Gold present the same probability distribution
 $P(x)$.}
\label{fig1}
\end{figure}

%=========================================================================
\subsection{Experimental Analysis of $P(x)[\Delta t]$}
%=========================================================================

In Fig. \ref{fig1}, we present the log-returns  $x$ (five minutes sampling) for a large set of futures,
with $\Delta t = 5$ minutes.
To make the comparison, we have scaled $x$ for all futures to the volatility of the DAX futures (FDAX).
All data sets cover the period $2001$-$2011$.
On the left hand side of Fig. \ref{fig1},
we observe that futures on DAX, Bund, Yen, Euro, Gold present the same probability distribution
for $x$. Therefore $P(x)$ is universal for all these data series once the volatility is normalized to the same value.
Note that there are futures on commodities which exhibit some larger tails \cite{b6} and we can not claim
the universality of this distribution whatever  futures. See also Ref. \cite{b5}.
%On the right hand side of Fig. \ref{fig1}, we provide comparisons with futures on commodities.
%We observe that data series on CL (Crude Oil) follows the same $P(x)$ as FDAX, but
%other data series on Wheat and NG (Natual Gaz) exhibit some larger tails.

\begin{figure}[htbp]
  \begin{center}
    \includegraphics[width=0.4\textwidth]{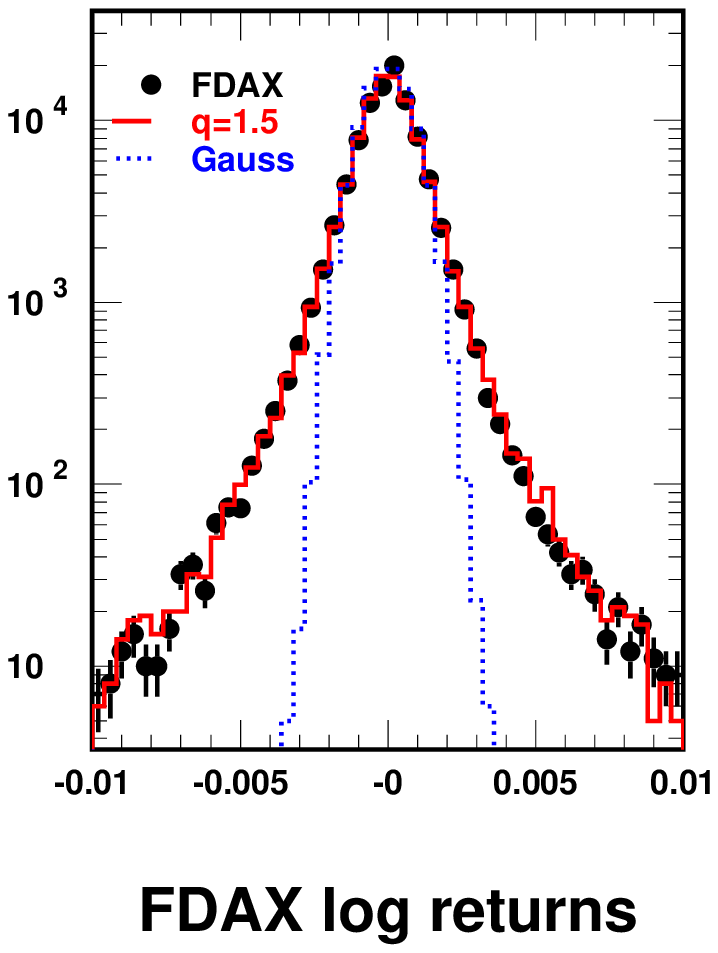}
    \includegraphics[width=0.4\textwidth]{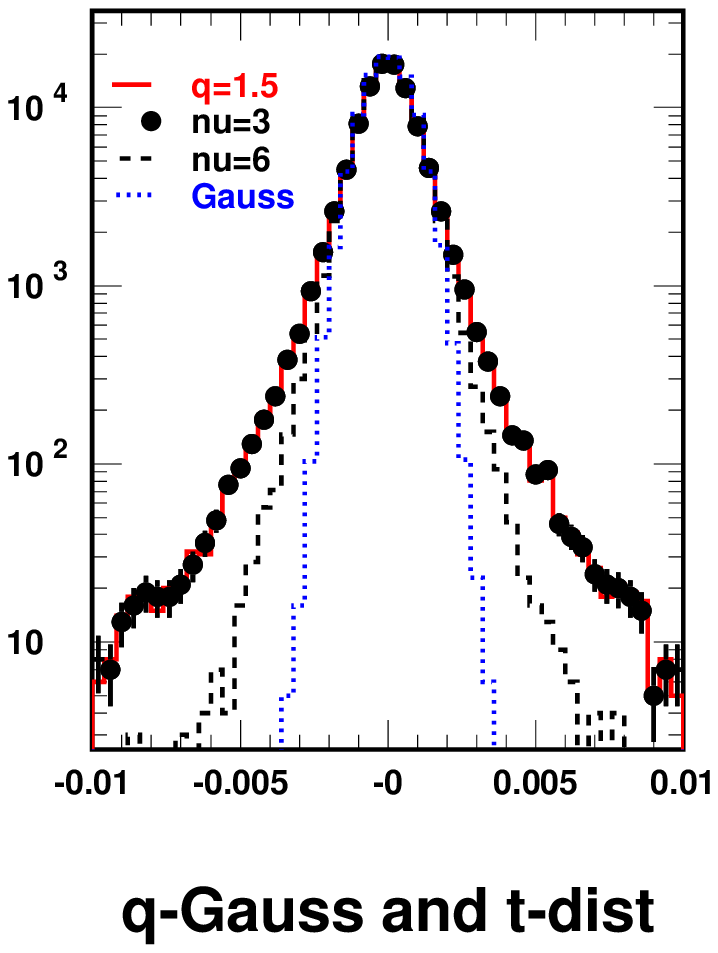}
  \end{center}
  \caption{
We show that the q-exponential probability
distribution, with $q=1.5$, gives a good description of the data (Right).
It corresponds to $\nu=3$ in the form of the t-distribution (Left).
The Gaussian approximation fails to describe properly the data.
When $\nu$ is increased, the Gaussian limit is approached.
}
\label{fig2}
\end{figure}

We can use results developed above in order to compare with experimental distributions $P(x)$
of Fig. \ref{fig1} (with fixed $\Delta t = 5$ minutes). We use predictions exposed in cases (i) and (iii), respectively
the Gaussian and the q-exponential forms (or t-distribution).
Results are shown in Fig. \ref{fig2}.
For data, we only display $P(x)$ for DAX futures.

In Fig. \ref{fig2}, we show that the q-exponential probability
distribution  of Eq. (\ref{q}), with $q=1.5$, gives a good description of the data.
Similarly, it corresponds to $\nu=3$ in the form of the t-distribution of Eq. (\ref{nu}).
Also, we observe  in Fig. \ref{fig2} that the Gaussian approximation fails to describe properly the data.
On the right hand side of Fig. \ref{fig2}, we observe that when $\nu$ is increased above $3$ in
 Eq. (\ref{nu}), then $P(x)$ stands in the middle of the correct probability density and the Gaussian
approximation. When $\nu$ tends to infinity, the t-distribution recovers the Gaussian limit.

In summary,
from the theory point of view, the probability distribution of log-returns of
financial time series, sampled at high frequency, 
can be expressed quite generally as $P(x)=\frac{1}{Z} \exp( - w(x)/D)$. 
In particular, we have shown that a large sample of high liquid futures 
(with normalized log-returns) are compatible with a distribution of the form
\be
\label{ff}
P(x) \propto \frac{1}{(1+\frac{x^2}{\nu})^{(\nu+1)/2}} \ \ \ , \ \ \ \nu \simeq 3.
\ee
Commodity futures deviate from this shape with larger tails and thus a smaller value of the exponent $\nu$.

\begin{figure}[htbp]
  \begin{center}
    \includegraphics[width=1\textwidth]{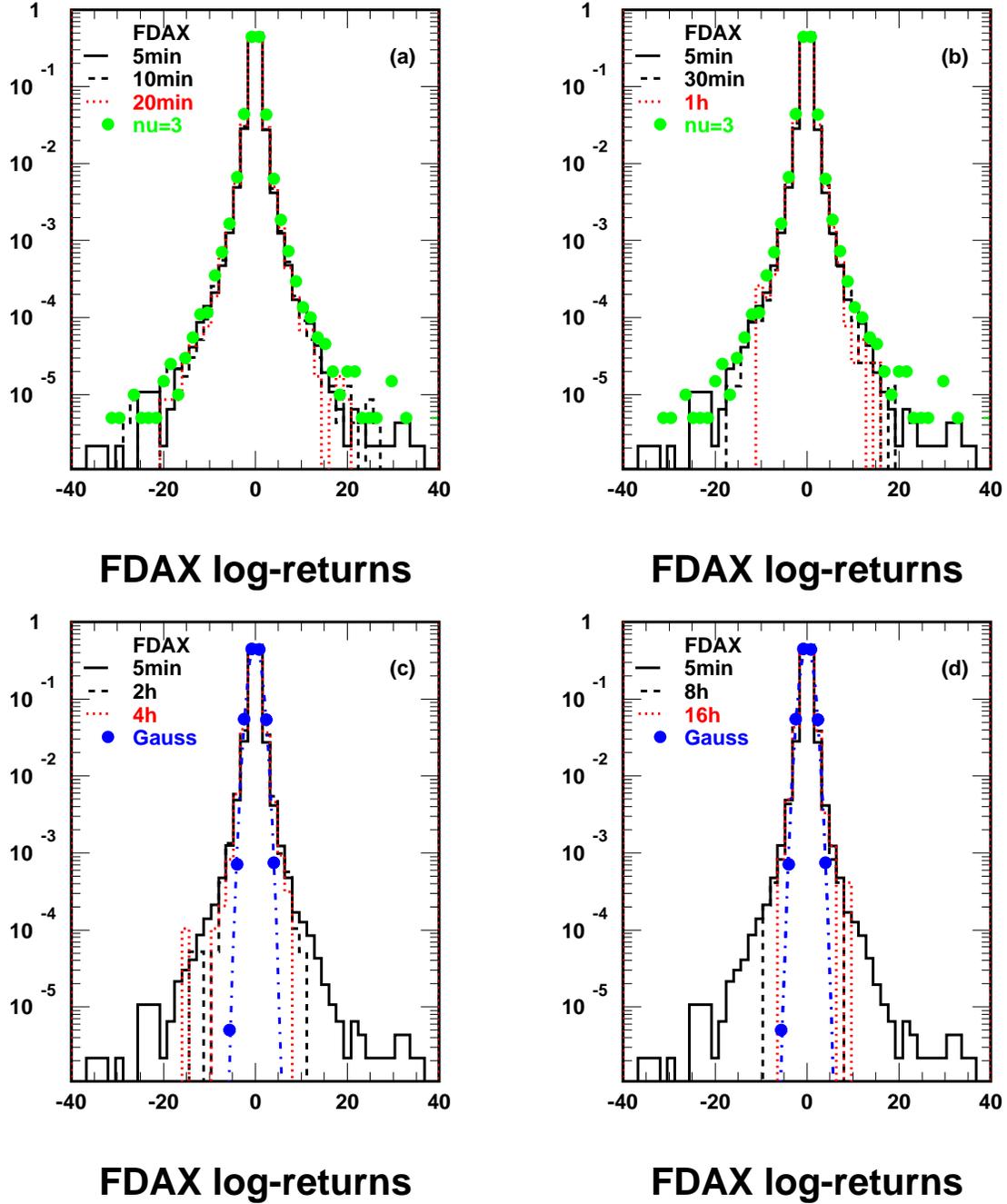}
  \end{center}
  \caption{
Normalized log-returns for the FDAX on different time scales 
$\Delta t$. The normalization is done by changing $x$ in $(x-\mu_x) / \sigma_x$ where $\mu_x$
and $\sigma_x$ are  the sample mean and variance. This re-definition does not change obviously the tail behavior of $P(x)$.
The basic resolution $\Delta t= 5$ minutes is displayed in all cases (a)-(d).
Then, we show the distributions using
  $\Delta t= 10,20$ minutes (a),
$\Delta t= 30,60$ minutes (b),  $\Delta t= 2,4$ hours (c)
and finally 
$\Delta t= 8,16$ hours  (d).
For small times scales $\Delta t < 1$ hour (a)-(b), we confirm that
the distributions of log-returns follows Eq. (\ref{ff}) with $\nu=3$.
When the resolution is increased to $\Delta t =2,4$ hours (c), we observe that we start approaching the 
Gaussian regime. This regime is reached for  $\Delta t \ge 8$ hours as can be seen on (d).
}
\label{fig3}
\end{figure}

\begin{figure}[htbp]
  \begin{center}
    \includegraphics[width=1\textwidth]{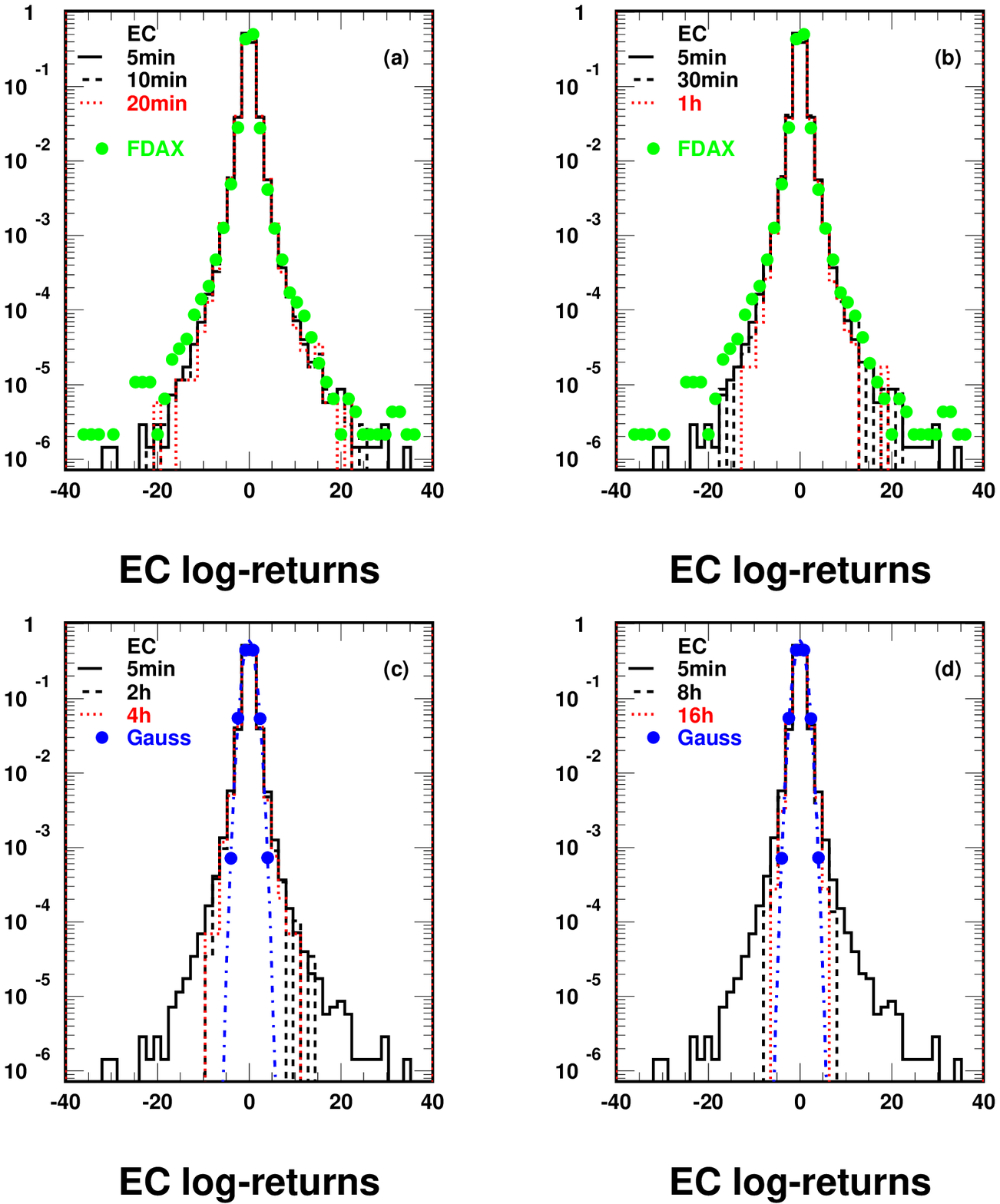}
  \end{center}
  \caption{
Normalized log-returns for the Euro futures (EC) on different time scales 
$\Delta t$. The normalization is done by changing $x$ in $(x-\mu_x) / \sigma_x$ where $\mu_x$
and $\sigma_x$ are  the sample mean and variance. This re-definition does not change obviously the tail behavior of $P(x)$.
The basic resolution $\Delta t= 5$ minutes is displayed in all cases (a)-(d).
Then, we show the distributions using
  $\Delta t= 10,20$ minutes (a),
$\Delta t= 30,60$ minutes (b),  $\Delta t= 2,4$ hours (c)
and finally 
$\Delta t= 8,16$ hours  (d).
For small times scales $\Delta t < 1$ hour (a)-(b), we confirm that
the distributions of log-returns follows the same distribution probability as the FDAX.
When the resolution is increased to $\Delta t =2,4$ hours (c), we observe that we start approaching the 
Gaussian regime. This regime is reached for  $\Delta t \ge 8$ hours as can be seen on (d).
}
\label{fig4}
\end{figure}

%%%%%%%%%%%%%%

Once Eq. (\ref{ff}) is established, 
we can examine the validity of this probability distribution as a function of $\Delta t$, from $5$ minutes to several hours.
This is done in Fig. \ref{fig3} for the DAX futures (FDAX) and Fig. \ref{fig4} for the Euro futures  (EC).
As mentioned above, these two data sets cover the period $2001$-$2011$.
In Fig. \ref{fig3} (a)-(d), we present the normalized log-returns for the FDAX on different 
time scales (or resolutions) 
$\Delta t$. The normalization is done by changing $x$ in $(x-\mu_x) / \sigma_x$ where $\mu_x$
and $\sigma_x$ are  the sample mean and variance. This re-definition does not change obviously the tail behavior of $P(x)$.

The basic resolution $\Delta t= 5$ minutes is displayed in all cases (a)-(d)
as a reference.
Then, we show the distributions using
  $\Delta t= 10,20$ minutes for Fig. \ref{fig3} (a),
$\Delta t= 30,60$ minutes for Fig. \ref{fig3} (b),  $\Delta t= 2,4$ hours  for Fig. \ref{fig3} (c)
and finally 
$\Delta t= 8,16$ hours  for Fig. \ref{fig3} (d).
For small times scales $\Delta t \le 1$ hour, in Fig. \ref{fig3} (a)-(b), we confirm that
the distributions of log-returns follows Eq. (\ref{ff}) with $\nu=3$.

When the resolution is increased to $\Delta t =2,4$ hours, Fig. \ref{fig3} (c), we observe that 
$P(x)[\Delta t]$ starts approaching the 
Gaussian regime. This regime is reached for  $\Delta t \ge 8$ hours as can be seen on \ref{fig3} (d).

In Fig. \ref{fig4} (a)-(d), we present similar plots for the Euro futures. Same conclusions can be derived.
For small times scales $\Delta t < 1$ hour, in Fig. \ref{fig4} (a)-(b), we observe that
the distributions of log-returns follows Eq. (\ref{ff}) with $\nu=3$. 
When the resolution is increased to $\Delta t =2,4$ hours, Fig. \ref{fig4} (c), we observe that 
$P(x)[\Delta t]$ starts approaching the 
Gaussian regime, which is reached for  $\Delta t \ge 8$ hours (\ref{fig4} (d)).

Identical conclusions are valid also for futures displayed in Fig. \ref{fig1}.
This is a  reasonable behavior which is compatible with the general statement done in the introduction.
$P(x)[\Delta t]$  evolves from
power law at small time scales $\Delta t$ to Gaussian at large time scales,
namely $\Delta t \ge 8$ hours. With the above analysis, we have put
quantitative estimates on this statement for several high liquid futures.

%%%%%%%%%%%%%%
%%%%%%%%%%%%%%

%=========================================================================
\section{Factorial Moment Analysis}
%=========================================================================
\label{facmom}

%=========================================================================
\subsection{Definitions}
%=========================================================================

In this section, we intend to analyze the dynamics of returns
using an alternative technique.
The main interest is to show the consistency of all results within different techniques.
In the following, we are interested in the multiplicity of
positive or negative returns in a given time window $\Delta t$.
Indeed, sequences of positive and negative returns are much indicative of the
statistical nature of fluctuations in the price series.
The idea is then to extract a quantitative information from these
sequences. 

First,  we present the  situation in
nuclear physics.
At nuclear or sub-nuclear energies,
the number of hadrons created during inelastic
collisions varies from one event to another. 
The distribution of the number of produced hadrons, namely
the multiplicity distribution, 
provides a basic means to characterize the events.
The multiplicity distribution contains information 
about multi-particle correlations in an integrated form,
providing a general and sensitive means to probe
the dynamics of the interaction.
Particle multiplicities  
can be studied in terms of the 
normalized factorial moments
\begin{equation}
\label{facm}
F_q(\Gamma)=\la n(n-1)\ldots (n-q+1)\ra / \la n \ra^q,
\qquad q=2,3, \ldots ,  
\end{equation}
for a specified phase-space region of size $\Gamma$. 
The number, $n$, of particles is measured inside  $\Gamma$
and angled brackets $\la\ldots\ra$ denote 
the average over all events. The factorial
moments
are convenient tools  to characterize the multiplicity
distributions when $\Gamma$ becomes small.
For uncorrelated particle production within $\Gamma$, 
Poisson or Gaussian  statistics
holds  and $F_q=1$ for all $q$. Correlations
between particles lead to a broadening of the multiplicity distribution 
and to dynamical fluctuations. In this case, the normalized factorial 
moments increase  
with decreasing $\Gamma$. 
The idea is then to divide the factorial moment defined in Eq. (\ref{facm})
in more and more bins. 

We can thus compute the related moment following
Eq. (\ref{facm}) as
\be
F_{i}=\frac{1}{N_{events}}\sum_{events}
\frac{\sum_{k=1}^{n_{bins}} \left\{n_{k}(n_{k}-1)
\cdots (n_{k}-i+1)\right\}/n_{bins}}
{(\langle n\rangle /n_{bins})^{i}}
\label{genfacmom}
\ee
where $\langle n \rangle$ is the average number of particles in the full
phase space region accepted ($\Gamma$), $n_{bins}$ denotes the number of bins in
this region and $n_k$ is the multiplicity in $k$-th bin.

The behavior of factorial moments plotted as a function of the number of bins 
(which means decreasing bin sizes)
provides information about the character of multiplicity
fluctuations among different bins. Rising of $F_i$ with rising $n_{bins}$
(decreasing bin size) generally signalizes deviation from purely
Gaussian distribution of fluctuations. The linear growth
of $\log F_i$ with $n_{bins}$ was defined as intermittency in \cite{BiaP}. 
See also \cite{Lipa,Dremin:1993dt,Dremin:1993ee,Rames:1994qm}. In the following, 
the term is  used for any
type of growth of $F_i$ observed.

As a matter of fact, it has  
been noticed in \cite{BiaP} that the use of  
factorial moments allows to extract the dynamical signal from the  
Poisson noise in the analysis of the multiplicity  
signal in high energy reactions. 

In addition, it has been shown that it is possible to define and compute a multi-fractal  
dimension, $\clDq$, for the theory of strong interactions \cite{DD93,OW93}
\begin{equation}
\label{inter}
F_q(\Gamma)=\la n(n-1)\ldots (n-q+1)\ra / \la n \ra^q \propto \Gamma^{(q-1)(1-{\cal D}_q/d)} 
\end{equation}
where d is the dimension of the phase space under consideration  
($d=2$ for the
whole angular phase space, and $d=1$ if one has integrated over, say  
the
azimuthal angle).   
In the constant coupling case $\clDq$ is well defined and
reads
\begin{equation}
\label{diminter}
{\cal D}_q = \gamma_0\frac{q+1}{q} 
\end{equation}
where $\gamma_0^2=4C_A{\alpha_S/ 2\pi}$, $\alpha_S$ is the strong interaction  
coupling constant, $C_A$ is the gluon color factor
\cite{DD93,OW93}.   
The choice of the factorial moments as a specific tool in order to study  
the scaling behavior of the high energy multiplicity  
distributions is then useful to analyze the underlying dynamics of the
processes. In principle, we can  extend this last idea to other fields where
factorial moments can be defined.

%=========================================================================
\subsection{Generating Function for Factorial Moments}
%=========================================================================
\label{formalism}

The multiplicity distribution is defined as
$
  P_{n} = {\sigma_{n}}/{\sum_{n=0}^{\infty}\sigma_{n}}    
$,
where $\sigma_{n}$ is the cross section of an
$n$-particle production process 
(the so-called topological cross section)
and the sum is over all possible values of $n$ so that
\begin{equation}
  \sum_{n=0}^{\infty}P_{n} = 1 .          \label{2}
\end{equation}
The  generating function can  be
defined as
\begin{equation}
  G(z) = \sum_{n=0}^{\infty }P_{n}(1+z)^{n}  ,    \label{3}
\end{equation}
which substitutes an analytic function of $z$ in place of the set of numbers 
$P_{n}$.
Then, we obtain the factorial moment or order $q$ as
\begin{equation}
  F_{q} = 
\frac {1}{\langle n \rangle ^{q}}\cdot \frac {d^{q}G(z)}{dz^{q}}\vline _{z=0}, 
\label{4}
\end{equation}
and the corresponding definition for cumulants
\begin{equation}
  K_{q} = \frac {1}{\langle n \rangle ^{q}}\cdot \frac {d^{q}\ln G(z)}{dz^{q}}.
\vline _{z=0}, \label{5}
\end{equation}
 
The expression for $G(z)$ can then be re-written as
\begin{equation}
  G(z) = \sum _{q=0}^{\infty } \frac {z^q}{q!} \langle n \rangle ^{q} F_{q} 
  \;\;\;\; ( F_0 = F_1 = 1 ),    \label{7}
\end{equation}
\begin{equation}
  \ln G(z) = \sum _{q=1}^{\infty } \frac {z^q}{q!} \langle n \rangle ^{q} K_{q} 
  \;\;\;\; ( K_1 = 1 ).    \label{8}
\end{equation}
The physical meaning of these moments has been discussed in
the previous section. Another interpretation can be seen 
from the above definitions if they are presented in the form of 
integrals of correlation functions. 
Let the single symbol $y$ represent all kinematic variables needed 
to specify the position of each particle in
the phase space volume~$\Gamma $ \cite{d1}. 
A sequence of inclusive 
$q$-particle differential cross sections $d^{q}\sigma /dy_{1}\ldots dy_{q}$
defines the factorial moments as
\begin{equation}
  F_{q} \sim \frac{1}{\langle n\rangle ^q}\int _{\Gamma }dy_{1} 
  \ldots \int _{\Gamma } dy_{q}\frac {d^{q}\sigma }{dy_{1}\ldots dy_{q}}. 
\label{12b}
\end{equation}
Therefore, factorial
moments include all correlations within the system of
particles under consideration. 
They represent integral characteristics of any correlation
between the particles whereas cumulants of rank $q$ represent genuine
$q$-particle correlations not reducible to the product of lower order
correlations.

%=========================================================================
\subsection{Experimental Analysis}
%=========================================================================

\begin{figure}[htbp]
  \begin{center}
    \includegraphics[width=0.4\textwidth]{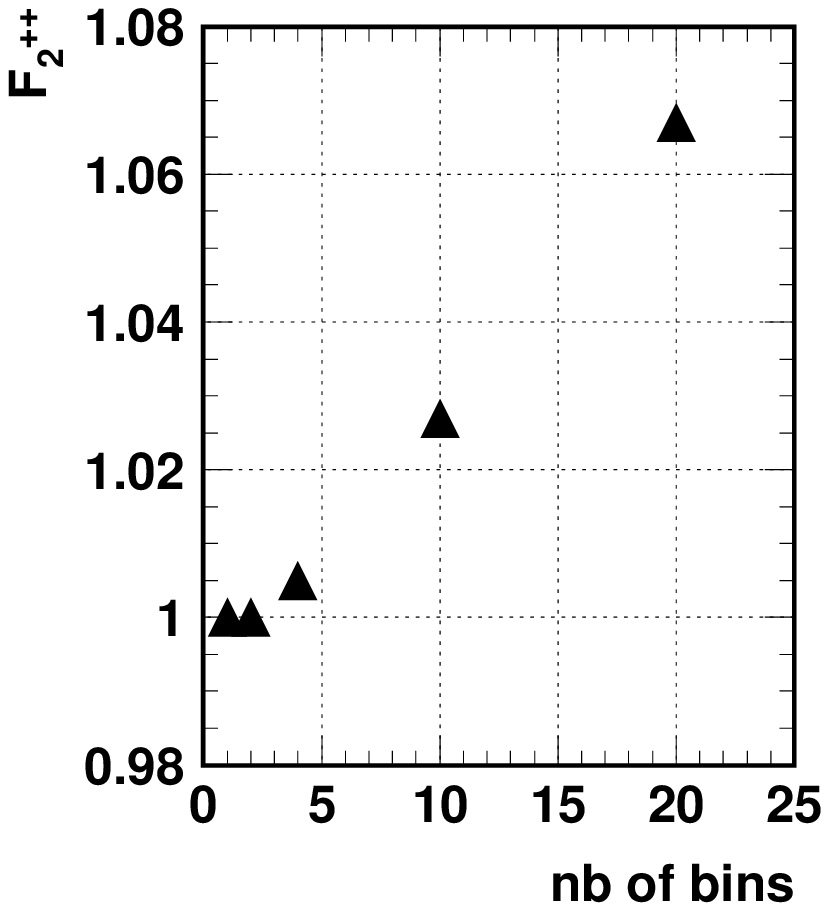}
    \includegraphics[width=0.4\textwidth]{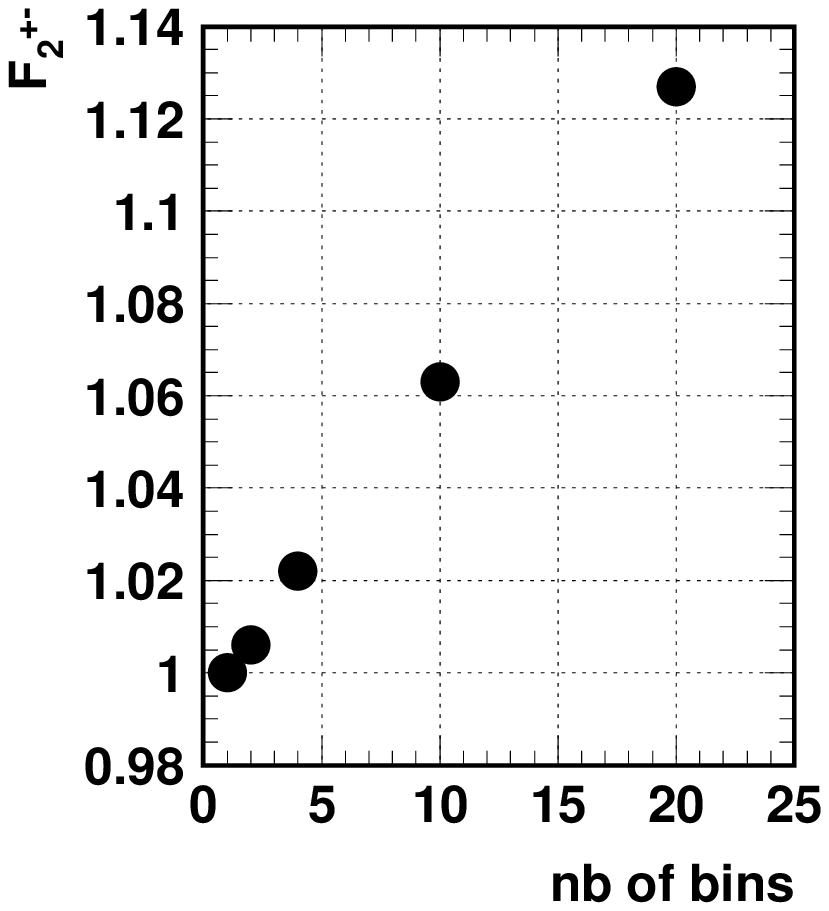}
  \end{center}
  \caption{Euro futures. 
Left: factorial moments $F_2^{++}$ are displayed for $1$, $2$, $4$, $10$ or $20$ bins.
The global time window of $16.5$ hours ($n_{bins}=1$).
This provides a time scale ranging from $1.6$ hours for $20$ bins till $16.5$ hours for $1$ bin.
We observe that for $1$ and $2$ bins segmentation, $F_2^{++}$ is found equal to $1$
and $F_2^{++}$ is increasing above $1$ when the number of bins gets larger than $4$.
This confirms that non-Gaussian fluctuations in the sequence of returns
returns start to play a role when the scale (resolution)
is below $4$ hours. 
Right: similar results and comments for unlike-sign factorial moments $F_2^{+-}$. 
Note that intermittency is enhanced 
for $F_2^{+-}$
compared to the like sign quantity $F_2^{++}$.}
\label{f2pp}
\end{figure}

The analogy with returns of financial price series is immediate.
If we divide the price series $y(t)$ in consecutive time windows of lengths $\Delta t$,
we define like this a set of events.
In each window, we have a certain number of positive returns  $n_+$,
where $y(t)-y(t-1) >0$, and similarly of
negative returns $n_-$.
If the sequence of returns is purely
 uncorrelated, following a  
Gaussian  statistics at all scales,
we expect $F_q=1$ for all $q$. 

However, correlations
between returns may lead to a broadening of the multiplicity distributions 
($n_+$ or $n_-$ or even a combination of both) 
and to dynamical fluctuations. In this case, the  factorial 
moments may increase  
with decreasing $\Delta t$, or increasing the number of bins that divide $\Delta t$, as in
Eq. (\ref{genfacmom}). In the following, we consider only the factorial moment of
second order $F_2$.
We can write
\be
F_{2}^{++}=\frac{1}{N_{events}}\sum_{events}
\frac{\sum_{k=1}^{n_{bins}} \left\{n_{k}^{+}(n_{k}^{+}-1)
\right\}/n_{bins}}
{(\langle n^{+}\rangle /n_{bins})^{2}}
\label{like}
%\end{eqn}
\ee
where $\langle n \rangle$ is the average number of positive returns in the full
time window  ($\Delta$), $n_{bins}$ denotes the number of bins in
this window and $n_k^{+}$ is the number of positive returns in $k$-th bin.

We consider the data  series on  the Euro futures, sampled in 5 minutes units
(same data set as in section \ref{returns}).
We present calculations for a time window 200 times units, that we
divide in $1$, $2$, $4$, $10$ or $20$ bins.
This means that the  time resolution extends from $1.6$ hours to  $16.5$ hours.
Note that with a time window of 200 time units, we set up an ensemble of more that 3400 events.
The statistical precision of the following analysis is then ensured.
Results are presented in Fig. \ref{f2pp} (left) for positive returns.
Factorial moments $F_2^{++}$ are displayed for $1$, $2$, $4$, $10$ or $20$ bins.
As mentioned above, this gives a time resolution ranging from $1.6$ hours for $20$ bins till $16.5$ hours for $1$ bin.
The statistical precision is of $0.1$\%. We can define a systematical uncertainty
by shifting the time window of $200$ units by $50$ or $100$ units, which means that 
we define a different set of events among the price series. Variations in the
calculations of $F_2^{++}$ are lower that $0.1$\%. Fig. \ref{f2pp} (left) displays the
full uncertainty of these quantities.

In Fig. \ref{f2pp} (left),
we observe that for $1$ and $2$ bins segmentation $F_2^{++}$ is found to be equal to $1$.
As expected, for the larger resolution, positive returns appear as completely uncorrelated.
When the number of bins is increased, we observe the phenomenon described in 
section 1, with an enhanced sensitivity of $F_2^{++}$ to non-Gaussian fluctuations.
This confirms that correlations
between positive returns start to play a role when the resolution
is below $4$ hours. 
Thus, Fig. \ref{f2pp} (left) exhibits a clear feature of intermittency.
As in nuclear reactions, an increase of the resolution
to a certain extend leads to a broadening of the multiplicity distribution 
and to super-diffusive fluctuations. Let us note that this is a feature that can be
approached in the context of non-extensive statistics \cite{ts1,ts2,ts3}.

Similar results can be obtained for $F_2^{--}$, defined for negative returns distribution.
\be
F_{2}^{--}=\frac{1}{N_{events}}\sum_{events}
\frac{\sum_{k=1}^{n_{bins}} \left\{n_{k}^{-}(n_{k}^{-}-1)
\right\}/n_{bins}}
{(\langle n^{-}\rangle /n_{bins})^{2}}
\label{like2}
%\end{eqn}
\ee
where $\langle n \rangle$ is the average number of negative returns in the full
time window  ($\Delta$), $n_{bins}$ denotes the number of bins in
this window and $n_k^{-}$ is the number of negative returns in $k$-th bin.
For all values displayed in Fig. \ref{f2pp} (left) for $F_2^{++}$, we derive the
same result for $F_2^{--}$ 
 up to $0.1$\%, which makes $F_2^{++}$ and $F_2^{--}$ 
indistinguishable. 

A direct extension of the above study can be
obtained if we examine moments for like-sign and
unlike-sign combinations of returns separately.
The like-sign factorial moment of order $2$ is defined by Eq. (\ref{like}).
The unlike-sign can be expressed as 
\be
F_{2}^{+-}=\frac{1}{N_{events}}\sum_{events}
\frac{\sum_{k=1}^{n_{bins}} \left\{n_{k}^{+}n_{k}^{-}
\right\}/n_{bins}}
{\langle n^{+}\rangle \langle n^{-}\rangle /(n_{bins})^{2}}
\label{unlike}
\ee
Results are presented in Fig. \ref{f2pp} (right) for $F_{2}^{+-}$.
Here again, we observe intermittency, with an increase of $F_{2}^{+-}$ 
as a function of the number of bins.
Also, this increase is larger than for like-sign calculations as can be observed when
comparing Fig. \ref{f2pp} (left) and \ref{f2pp} (right).
The sensitivity to non-Gaussian fluctuations in the returns sequence 
is thus enhanced with the definition of d to $F_{2}^{+-}$. This is also an effect
observed in nuclear interactions \cite{Dremin:1993dt,Dremin:1993ee,Rames:1994qm}.

\begin{figure}[htbp]
  \begin{center}
    \includegraphics[width=0.4\textwidth]{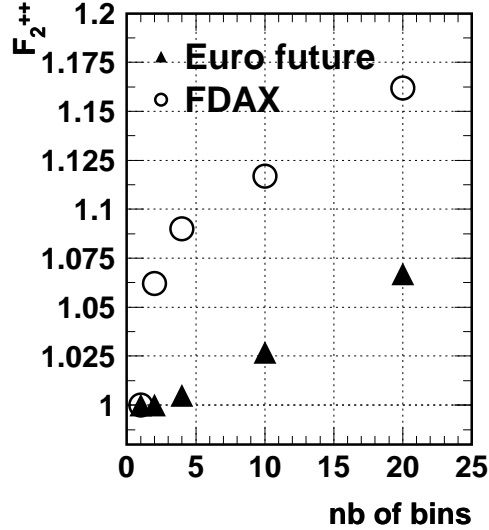}
  \end{center}
  \caption{Euro and DAX futures. Factorial moments $F_2^{++}$ are displayed for $1$, $2$, $4$, $10$ or $20$ bins.
The global time window of $16.5$ hours ($n_{bins}=1$).
This provides a time scale ranging from $1.6$ hours for $20$ bins till $16.5$ hours for $1$ bin.
For the DAX future, the intermittent behavior is more pronounced than for the Euro. }
\label{f2pp2}
\end{figure}

\begin{figure}[htbp]
 \begin{center}
   \includegraphics[width=0.4\textwidth]{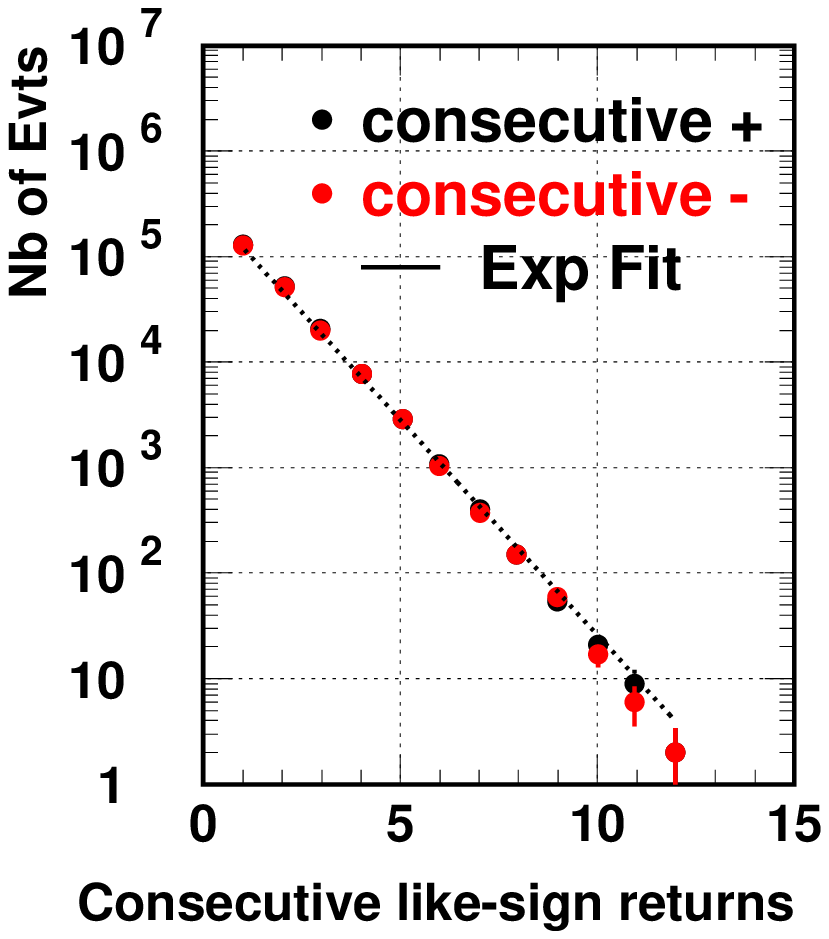}
   \includegraphics[width=0.4\textwidth]{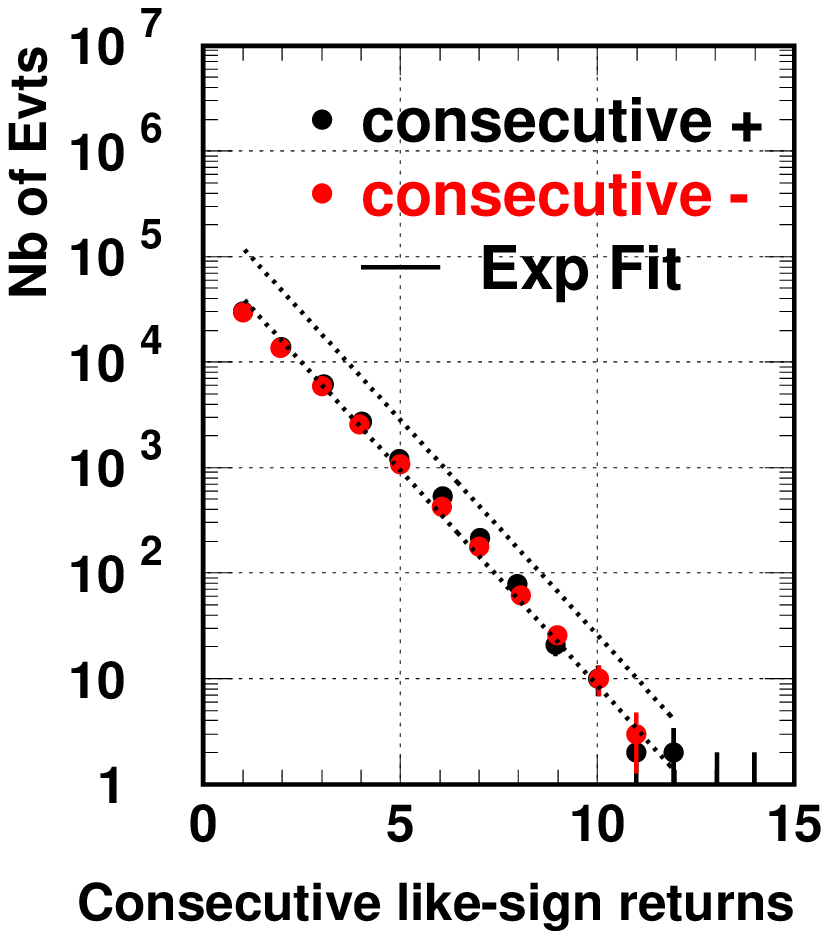}
 \end{center}
\caption{Left: Euro futures sampled in 5min time units.
Right: Euro futures sampled in 30min time units.
The distribution of events (probability distribution) is displayed as a function of the
size of the gap.
The gap is defined as the number of consecutive positive returns after a negative return,
thus this is a gap in negative return. Inversely for a gap in positive return.
The Euro futures
 series over 10 years is sampled in two different time units (5min-Left and 30min-Right).
In both cases, we observe effectively an exponential fall of the probability distribution
as a function of the gap size. 
We have presented an exponential fit on top of each plot. On the Right plot (30 minutes sampling),
we super-impose also the fit obtained on the 5 minutes sampling, in order to show that
the exponential slope is identical in both cases.
}
\label{gap}
\end{figure}

A final comment is in order. This analysis has been illustrated on the Euro futures.
However,
we have found that a similar intermittent behavior is observed on other futures, 
like the DAX futures (FDAX).
For each series, the values of the like-sign and unlike-sign factorial moments of order $2$
vary compared to results exhibited for the Euro future.
However the rise of the $F_2$ as a function of the number of bins is an invariant
property, with always $F_2^{++}$ and $F_2^{+-}$ equal to unity for a time window of
order $16$ hours. For some series, like FDAX, the intermittent behavior already prevails for
a time scale of order $8$ hours, with an increase of $F_2^{++}$ from $1$ ($n_{bins}=1$, $\Delta t\sim16$~h)
to $1.16$ ($n_{bins}=20$, $\Delta t\sim1.6$~h).
This is shown in Fig. \ref{f2pp2}.
This growth with $n_{bins}$ is faster than the Euro future. However, the key feature is always the same.
For a sufficiently fine time scale, more precisely for  $\Delta t$ smaller than $4$ to $8$ hours
depending of the financial product, intermittency  is observed.

Interestingly,
we can use this formalism 
in order to derive some further statements on correlations of returns. From Eq. (\ref{7}),
we get
$$
G(-1)=p_0
$$
which corresponds to the probability to have 
zero positive (resp. negative) returns in a given time window.
This defines a gap in duration for 
positive (resp. negative) returns.
Let us use Eq. (\ref{5}) to express $G(-1)$ in another way
using cumulants $K_q$
$$
G(-1)=\exp(-K_1 +K_2/2! +...)
$$
When we can neglect correlations within a large time window,
we have shown $K_2=0$, then we conclude
\be
p_0 \simeq \exp(-K_1) = \exp(-<n_{+,-}>) = \exp(-\rho \Delta t)
\label{final}
\ee
where $\Delta t$ represents the time scale (time window). 
This last expression is very simple and instructive. It states that the  gap
probability is exponentially suppressed with this time scale.
This is illustrated in Fig. \ref{gap} for the finance case.
We observe the distribution of events (probability distribution) as a function of the
size of the gap. 
The gap is defined as the number of consecutive positive returns after a negative return,
thus this is a gap in negative return. Inversely for a gap in positive return.
This gap is given in number of time units for the financial time series
considered. In Fig. \ref{gap}, we display results for  the Euro futures
 series over 10 years, sampled in two different time units.
In both cases we observe effectively an exponential fall of the probability distribution
as a function of the gap size.  As illustrated also in Fig. \ref{gap},
this exponential fall does not depend on the sampling.
This confirms the prediction of Eq. (\ref{final}).

%=========================================================================
\section{Further Analysis of Correlations}
%=========================================================================

With factorial moments, we have discussed how the  
deviation from $F_q=1$ with the time scale  $\Delta t$
 is associated to a broadening of the probability distribution 
of returns
and then to dynamical fluctuations. 

Correlations
between returns can be tested  in  other ways. 
One general methodology consists in
estimating how a certain fluctuation measurement,labeled
generically as $F$, scales with the size $\tau$ of the time window
considered. Specific methods, such as the Hurst rescaled range
analysis \cite{hurst} or the Detrended Fluctuation Analysis
\cite{stanley1,jafferson}, differ basically on the choice of $F(\tau)$.  If the
financial time series is uncorrelated one expects that $F\sim\tau^{1/2}$, as is
the case for the standard Brownian motion. On the other hand, if
$F\sim\tau^H$ with $H\ne1/2$ one then says that the time series has
long-term memory. The exponent $H$ is generally
referred to as the Hurst exponent.

To understand simply the interest of this exponent in describing correlations, 
we recall  that the fractional Brownian motion (FBM) $\{B_H(t), t>0\}$ is a
Gaussian process with zero mean and stationary increments whose
variance and covariance can be written \cite{a1,a2,a3,a4}
\begin{eqnarray}
&&<B_H^2(t)>_t=t^{2H}, \label{eq:varBH}\\
&&<B_H(s)B_H(t)>_t=\frac{1}{2}\left(s^{2H}+
t^{2H}-|t-s|^{2H}\right), \label{eq:corBH}
\end{eqnarray}
where $0<H<1$ and $<\cdot>_t$ denotes expected value.

For $H=1/2$ the process $B_H(t)$ corresponds to the standard
Brownian motion, in which case the increments $X(t)=B_H(t+1)-B_H(t)$
are statistically independent and represent the usual white
noise. For $H \ne 1/2$ the increments $X_t$ display long-range correlation with
\begin{equation}
<[X(t+\tau)X(t)]>_t \simeq 2H(2H-1)\tau^{2H-2} \quad \mbox{for} \quad \tau\to\infty.
\end{equation}
where the average $<.>_t$ is made over $t$ and the full time window of the analysis.
Thus, if $1/2<H<1$ the increments of the FBM are positively correlated
and we say that the process $B_H(t)$ exhibits persistence. Likewise,
for $0<H<1/2$ the increments are negatively correlated and the FBM is
said to show anti-persistence \cite{a1,a2,a3,a4}.

%%%%%%%%%%%%%%%%
%%%%%%%%%%%%%%%% discussion
%%%%%%%%%%%%%%%%

Let us apply the discussion above on FBM $B_H(t)$ to real price series
presented in previous sections.
It is well known that the Hurst exponent can be extracted from financial price series using 
various techniques \cite{a1,a2,a3,a4}. In doing so, we need to compute $H$ from 
\be
\label{hh}
<x(t,\tau)^2>_t = < \left[\log [S(t+\tau)/S(t)] \right]^2>_t \propto \tau^{2 H}
\ee
in the range of validity of this expression.
Obviously, we can write
\be
\label{hh2}
x(t,\tau) = \sum_{t'=t}^{t'=t+\tau} x(t')
\ee

%Then, in a thirst part, we discuss the scaling of $<x(t,\tau)^2>_t$ with $\tau$ and relates the so-called Hurst exponent to the 
%power law tail.
At this level, we can safely assume that the returns $x(t')$ is identically distributed. However, the returns $x(t')$ 
are not independent and thus the central limit theorem does not apply. In order to lead intuition, we shall anyhow make the approximation that we can
sum up the sequence of $x(t')$ in Eq. (\ref{hh2}) using the central limit theorem. Then, there are two cases
\cite{a1,a2,a3,a4}:
 (i) if the power tail of the distribution of returns follows
$\nu>2$, the variance is finite and $x(t,\tau)$ converges to a Gaussian distribution; (ii) if $\nu<2$, the sum in Eq. (\ref{hh2})
tends to Levy stable distributions.
As we have shown in previous sections for the futures under consideration in this letter, we stand in the case (i).
Therefore, following the discussion above, we expect a Hurst exponent of order $1/2$.

\begin{figure}[htbp]
 \begin{center}
   \includegraphics[width=0.4\textwidth]{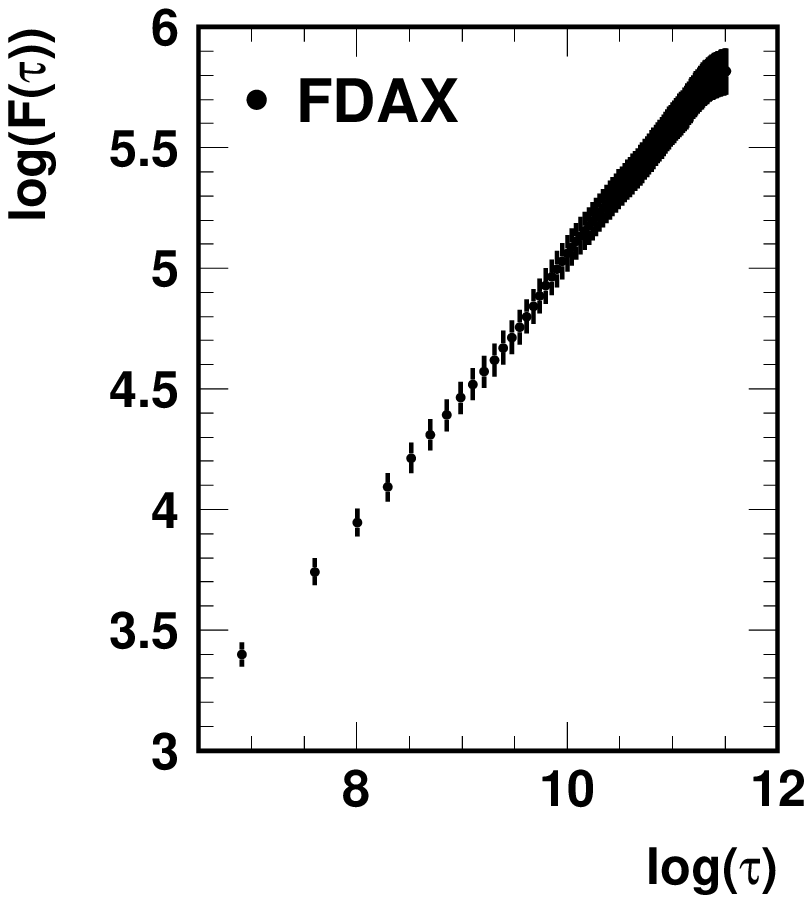}
   \includegraphics[width=0.4\textwidth]{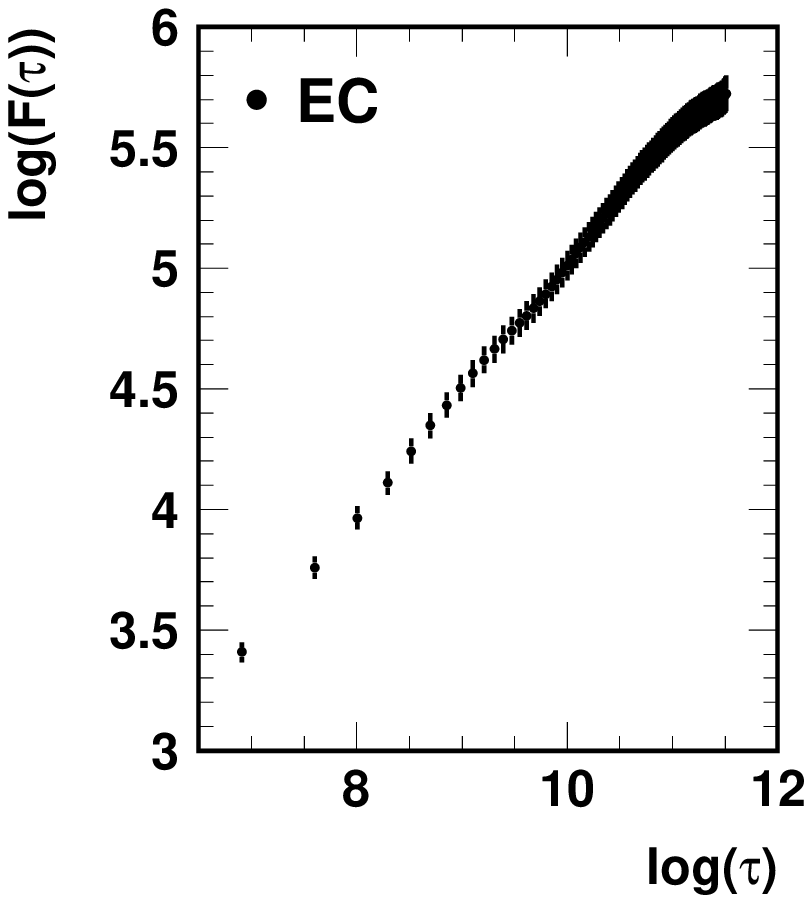}
 \end{center}
\caption{Left: DAX futures.
Right: Euro futures.
In these plots
$
F(\tau)=\frac{\sqrt{<|x(t+\tau)-x(t)|^2>_t}}{\sqrt{<|x(t)|^2>_t}}.
$
The Hurst exponent is extracted with the relation $F(\tau) \propto \tau^{H}$.
We get: $H=0.54 \pm 0.04$ for FDAX and $H=0.51 \pm 0.03$ for EC.
See text for a discussion on the uncertainties. 
}
\label{hhh}
\end{figure}

In order to determine the Hurst exponent experimentally, we compute
\be
\label{hform}
F(\tau)=\frac{\sqrt{<|x(t+\tau)-x(t)|^2>_t}}{\sqrt{<|x(t)|^2>_t}}.
\ee
Results are presented in
Fig. \ref{hhh} for the DAX and Euro futures.
Then, the Hurst exponent is extracted with the relation $F(\tau) \propto \tau^{H}$.
We get: $H=0.54 \pm 0.04$ for FDAX and $H=0.51 \pm 0.03$ for EC. These results are well compatible with the $1/2$ value
expected from the above discussion. They are also robust within the set of high liquid futures discussed in this letter.
Let us note that the uncertainties on $H$ are dominated by systematical effects. We have varied the time window for the calculation, the time interval and we have also divided the original data sets over 10 years on periods of 3 years.
Also, we have modified the definition of the time average defined in Eq. (\ref{hform}) using an exponential
smoothing in order to take into account the fact that the recent past is more important than the remote past.
The variations in the determination of the Hurst exponents give the errors quoted above.

Eq.  (\ref{hform}) can be extended to higher moments of the distribution of returns as
\be
\label{hformq}
F_q(\tau)=\frac{\sqrt{<|x(t+\tau)-x(t)|^q>_t}}{\sqrt{<|x(t)|^q>_t}},
\ee
which leads to the definition of generalized Hurst exponent, namely
$F_q(\tau) \propto \tau^{H(q) q/2}$.
Then, processes with $H(q)=H$ independent of $q$ are called uni-fractal whereas processes where
$H(q)$ is not constant are referred to as multi-fractal \cite{a1,a2,a3,a4}.
When considering all systematic effects listed above, we do not get any experimental sensitivity to the $q$ dependence 
in Eq. (\ref{hformq}).

%=========================================================================
\section{Conclusion and Outlook}
%=========================================================================

We have presented experimental results based on a large set of
time series on futures. When studying the probability distribution of log-returns,
we have shown that a t-distribution with $\nu \simeq 3$ gives a nice description
of almost all data series for a time scale $\Delta t$ below $1$ hour.
For $\Delta t \ge 8$ hours, the Gaussian regime is reached.
A particular focus has been put on the DAX and Euro futures, which both verify these
properties and thus exhibit a similar statistical texture.
This appears to be
a quite general result that stays robust on a large set of futures, but not on any data sets.
In this sense, this is not universal.

A technique using factorial moments defined on returns is described and similar results are obtained.
Let us recall that factorial
moments
are convenient tools  in nuclear physics to characterize the multiplicity
distributions when phase-space resolution becomes small.
In particular,
correlations
between particles lead to a broadening of the multiplicity distribution
and to dynamical fluctuations. In this case, the  factorial
moments increase  above $1$
with decreasing resolution. This corresponds to what can be called
intermittency.
A similar property has been illustrated on financial price series.
An intermittent behavior has been extracted
for DAX and Euro futures,
using moments of order $2$ ($F_2$). This
leads to perfectly consistent results with the standard distribution analysis.

Let us note that from a fundamental point of view, there is no clear reason why DAX and Euro futures
should present similar behavior with respect to their return distributions.
Both are complex markets 
where many  internal and external factors interact at each instant to determine the transaction price.
These factors are certainly different for an index on a change parity (Euro) and an index on stocks (DAX).
Thus, this is striking that we can identify universal statistical features in price fluctuations
of these markets. This is really the advantage of micro-structure analysis to prompt unified approaches 
of different kinds of markets. 

Finally, we have discussed that the power law distribution of returns is
well consistent with the information encoded into the Hurst exponent.
We have obtained
$H=0.54 \pm 0.04$ for DAX and $H=0.51 \pm 0.03$ for Euro futures.

An immediate outlook concerns the use of this knowledge to unfold the sequence of returns into
a more controlled sequence of trades. In Ref. \cite{lolo},  we have proven that it is possible to design 
such unfolding procedure. Interestingly, the trade durations derived on the DAX and Euro futures \cite{lolo}
correspond perfectly to the time scales derived in this letter. 
In fact, building a trading strategy is another consistency test of the knowledge we can get from the time series analysis.
The broadening of the probability distribution of returns when decreasing the time scale and the associated
dynamical fluctuations, as extracted from the factorial
moments  can serve as a guide in the unfolding procedure.

%=========================================================================

\end{document}